\magnification=1000
\hsize=12.6 true cm
\vsize=19.5 true cm
\baselineskip=14 pt
\hoffset=1.5 true cm
\voffset=2.0 true cm
\def\ref{\par\noindent\hangindent=0.75 true cm}

\input psfig.sty
\centerline{\bf A COMPARISON OF THE TIME VARIATION}
\centerline{\bf  OF RADIAL ABUNDANCE GRADIENTS}
\centerline{\bf FROM PLANETARY NEBULAE AND OPEN CLUSTERS}
\bigskip
\centerline{\it Walter  J. Maciel, R. D. D. Costa, P. A. L. Ferreira}
\smallskip
\centerline{IAG-USP}
\centerline{Rua do Mat\~ao 1226, Cidade Universit\'aria,
05508-900 S\~ao Paulo SP, Brazil} 
\centerline{maciel@astro.iag.usp.br}
\bigskip
\bigskip\noindent
{\baselineskip 10 pt {\bf Abstract.} 
The temporal behaviour of the radial abundance gradients has important 
consequences on models for the chemical evolution of the Galaxy. We present a 
comparison of the time variation of the abundance gradients in the Milky Way disk as 
determined from a sample of planetary nebulae and open clusters. We conclude that the 
[Fe/H] gradients as measured in open cluster stars strongly support the time flattening of 
the abundance gradient as recently determined from O/H measurements in planetary 
nebulae. We estimate the average flattening rate in the time interval 10 Myr $< \tau < $ 5 Gyr 
as roughly 0.005 dex kpc$^{-1}$ Gyr$^{-1}$.}
\bigskip
\bigskip\noindent
{\bf 1.  Introduction}
\bigskip\noindent
It is now generally accepted that the radial abundance gradients observed in the Milky Way 
disk are among the main constraints of models for the chemical evolution of the Galaxy. The 
study of the gradients comprise the determination of their average magnitudes along the disk, 
which include possible space variations, and their time evolution during the lifetime of the 
Galaxy (see for example Henry \& Worthey 1999, Maciel 2000 and Maciel \& Costa  2003 for 
recent reviews).

The magnitudes of the gradients can be derived from a variety of  objects, such as HII regions, 
early type stars, planetary nebulae, etc.  Recent investigations include Deharveng et al. 
(2000), Andrievsky et al. (2002) and Maciel et al. (2003), for HII regions, cepheid variables 
and planetary nebulae, respectively. Average values are generally in the range $-0.04$ to $-0.07$ 
dex/kpc  for the O/H gradient, with similar values for other element ratios  when available, as 
in the case of cepheids and planetary nebulae. 

The space variations of the gradients are more controversial. Some  flattening at large 
galactocentric distances is clearly discernible in a sample of galactic planetary nebulae, as 
shown by Maciel \& Quireza (1999) and more recently by Costa et al. (2004), on the basis of a 
study of nebulae located in the direction of the galactic anticentre. These results are supported 
by some  work on HII regions (see for example V\'\i lchez \& Esteban 1996). The cepheid  data 
are consistent with a flattened gradient  near the solar neighbourhood, as suggested by 
Andrievsky et al. (2002). On the other hand, apparently no flattening is observed in some 
recent studies of O, B stars in the galactic disk (see for example Smartt 2000).

Probably the most interesting property of the gradients is their time variation, as it appears to 
be a very distinctive constraint of many recent chemical evolution models (cf. Tosi 2003). As 
an example, models by Hou et al. (2000) and Alibés et al. (2001) predict a continuous time 
flattening of the gradients,  while models by Chiappini et al. (2001) are consistent with some  
steepening in a timescale of 3 to 10 Gyr. Therefore, it is extremely important to produce 
observational constraints to the time evolution of  the gradients, along with their magnitudes 
and possible space variations. Recently, Maciel et al. (2003) suggested that the O/H gradient 
is flattening from roughly $-0.11$ dex/kpc to $-0.06$ dex/kpc during the last 9 Gyr, or from $-0.08$ 
dex/kpc to $-0.06$ dex/kpc in the last 5 Gyr. These results were obtained using a large sample 
of planetary nebulae for which accurate abundances have been obtained, and for which the 
ages of the progenitor stars have been individually determined. As discussed by Maciel et al. 
(2003), the absolute ages derived are probably not accurate, but the relative ages of the stars 
are better determined, so that the time behaviour of the gradient can be derived, at least for the 
last 5 Gyr, which include most objects in the sample.

In this work, we compare the results obtained by Maciel et al.  (2003) for the planetary 
nebulae with some recent and  independent determinations based on open cluster stars. Along 
with planetary nebulae, open clusters are favorite objects to study the  time evolution of 
abundance gradients, as they comprise a wide age  bracket and have relatively well 
determined ages, based on a  detailed comparison of theoretical isochrones and color 
magnitude diagrams (see for example Friel 1999 and Phelps 2000 for recent reviews). The 
distances are also generally well determined, while the stellar metallicities, mostly derived by 
photometric techniques, are not as accurate as in the case of some elements in photoionized  
nebulae, but nevertheless the gradients can be derived within  a similar uncertainty, roughly $0.01$
dex/kpc. 
\bigskip
\bigskip\noindent
{\bf 2.  O/H gradients from planetary nebulae}
\bigskip\noindent
An estimate of the time variation of the O/H radial gradient in the galactic disk has recently 
been made by Maciel et al. (2003). From the observed O/H abundances in a large sample of 
nebulae, the [Fe/H] metallicity was determined on the basis of a correlation derived for disk 
stars. An age-metallicity relation was used to estimate the ages of the progenitor stars, so that 
the temporal behaviour of the gradients could be derived. Two cases were considered (A and 
B), in which the sample was divided  into three age groups, namely,  Case A: Group I, with 
ages in the range $0 < \tau$ (Gyr) $< 3$, Group II, for which $3 < \tau < 6$, and Group III, 
with $\tau  > 6 $ Gyr; and Case B: Group I, with ages in the range $0 < \tau  < 4$, Group II, 
for which $4 < \tau  < 5$, and Group III, with $\tau  > 5$  Gyr. 
The O/H gradients  (dex/kpc) derived by Maciel et al. (2003) are 
shown  in the third column of Table 1. Typical uncertainties of these gradients are of the order 
of 0.01 dex/kpc. For details on the determination of the gradients the reader is referred to the 
original paper. It can be seen that the O/H gradient flattens out from the older groups to the 
younger ones, which is especially true for Case B, in which the groups were chosen to have  
approximately the same number of objects. Although the choice of the age groups is arbitrary, 
in all cases the flattening of the gradient is apparent. As discussed by Maciel et al. (2003), 
even though the absolute  ages may be in error, it is unlikely that the relative ages are 
incorrect, so that the flattening of the gradient, although small, is probably real. In addition, 
Maciel et al. (2003) performed an independent calculation of the ages, on the basis of a 
relationship between the N/O abundance and the progenitor star mass, which led to the 
original stellar mass on the main sequence and therefore to another estimate of the age, with 
results similar to those shown in Table 1.
\bigskip\bigskip
\centerline{Table 1 – Abundance gradients: planetary nebulae}
$$\vbox{\halign{#\hfil&\qquad\hfil#\hfil&\qquad\hfil#\hfil
&\qquad\hfil#\hfil\cr\noalign{\hrule}
\ \cr
Group & Age (Gyr) & $d\log$(O/H)$/dR$ & $d$[Fe/H]$/dR$ \cr
\ \cr
\noalign{\hrule}
\ \cr
Case A &         &  &  \cr
I      & $0 - 3$ & $-0.065$ & $-0.079$ \cr
II     & $3 - 6$ & $-0.072$ & $-0.087$ \cr
III    & $> 6$   & $-0.116$ & $-0.141$ \cr
       &         &          &          \cr
Case B &         &          &          \cr
I      & $0 - 5$ & $-0.047$ & $-0.057$ \cr
II     & $4 - 5$ & $-0.089$ & $-0.108$ \cr
III    & $> 5$   & $-0.094$ & $-0.114$ \cr
\ \cr
\noalign{\hrule}}}$$
\bigskip
\bigskip\noindent
{\bf 3. [Fe/H] gradients from open clusters}
\bigskip\noindent
As mentioned in the Introduction, open clusters are favorite objects for which the 
determination of abundance gradients is possible (see for example Friel 1995, 1999,  and 
Phelps 2000 for recent reviews). Recently, new catalogues of open clusters have become 
available, which include space and kinematical data, as well as metallicities and estimates of 
the ages. In the work by Chen et al. (2003), which is based on the catalogue by Dias et al. 
(2002), a total of 119 clusters have been assembled for which distances and metallicities are 
available, which led to an average gradient of $d$[Fe/H]$/dR \simeq -0.063$ dex/kpc for the whole 
sample. This value is in good agreement with the average gradient of $-0.06$ dex/kpc obtained 
by Friel et al. (2002), based on an updated abundance calibration of spectroscopic indices in a 
sample containing 459 stars in 39 clusters. Taking into account two different age groups (ages 
$< 0.8$ Gyr and $\geq 0.8$ Gyr, respectively),  Chen et al. (2003) concluded that the iron gradient  
was steeper in the past, in agreement  with the results derived from O/H abundances in 
planetary nebulae. The conclusion by Chen et al. (2003) on the time variation of the  gradients 
confirms the earlier one by Friel et al. (2002), which is extremely important, as they have  
largely used different samples with similar results, so that the effect of a possible non-
homogeneity of the samples is probably small.
\bigskip\bigskip
\centerline{Table 2 - Abundance gradients: open clusters}
$$\vbox{\halign{#\hfil&\qquad\hfil#\hfil&\qquad\hfil#\hfil&\qquad\hfil#\hfil
&\qquad\hfil#\hfil\cr\noalign{\hrule}
\ \cr
Group & Age (Gyr) &  $d$[Fe/H]$/dR$ & $n$ & $r$ \cr
\ \cr
\noalign{\hrule}
\ \cr
Case A &         &   &  \cr
I      & $\leq 0.8$  & $-0.024$ & 80 & $-0.22$ \cr
II     & $> 0.8$     & $-0.075$ & 38 & $-0.70$ \cr
       &             &          &    &         \cr
Case B &             &          &    &         \cr
I      & $\leq 0.8$  & $-0.024$ & 80 & $-0.22$ \cr
II     & $0.8 - 2.0$ & $-0.067$ & 18 & $-0.69$ \cr
III    & $> 2.0$     & $-0.084$ & 20 & $-0.71$ \cr
\ \cr
\noalign{\hrule}}}$$
\bigskip
In order to obtain a more accurate comparison between the [Fe/H] gradients from open 
clusters and the planetary nebula data, we have used the sample by Chen et al. (2003) and 
considered two different cases, namely: Case A, with two age groups, and Case B, with 3 age 
groups, as defined in Table 2. We have assumed $R_0 = 7.6$ kpc for the distance of the LSR to 
the galactic center, as in Maciel et al. (2003), but the results are essentially unchanged if 
$R_0 = 8.0$ kpc or $R_0 = 8.5$ kpc. We have restricted our analysis to galactocentric distances 
$R < 16$ kpc, and excluded the object Berkeley 29, which does not satisfy this criterium (see Chen et 
al. 2003 for details). The total number of clusters in the sample is then 118. The derived 
[Fe/H] gradients (dex/kpc), the number of objects in each group ($n$) and the correlation 
coefficent ($r$) are shown in columns 3 to 5 of Table 2. The derived uncertainties in the 
gradients are in the range 0.01 to 0.02 dex/kpc. All age groups have galactocentric distances 
roughly in the range $R \simeq 6$ to 15 kpc, so that the derived gradients are representative of the 
galactic disk. There is some tendency for the clusters located at $R < 8$ kpc to be young, 
reflecting  the fact that older clusters are probably destroyed by collisions  with molecular 
clouds in the inner Galaxy.
\bigskip\noindent
{\bf 4. Discussion}
\bigskip\noindent
An analysis of the O/H gradients from planetary nebulae shown in Table 1 and the [Fe/H] 
gradients from open clusters (Table 2) confirm that the average gradients are flattening out for 
the last 8 Gyr approximately, as suggested by Chen et  al. (2003). We can improve on this 
conclusion by converting the O/H gradient into a [Fe/H] gradient, so that both sources of data 
can be directly compared. In fact, the O/H and [Fe/H] are supposed to be similar, but not 
exactly equal, which can be concluded by an inspection of the relation between the [O/Fe] 
ratio and  the metallicity [Fe/H] in the galactic disk. 

[Fe/H] gradients cannot be determined directly from planetary nebulae, as the iron lines are 
weak and a sizable fraction of this element is probably locked up in grains. However,  a 
relation between the iron and oxygen abundances can be  derived from observed properties of 
the stellar populations  in the galactic disk. A detailed discussion on the metallicities  and 
radial gradients from a variety of sources such as HII  regions, hot stars and planetary nebulae 
in the galactic disk  has been recently presented by Maciel (2002). According to this analysis, 
an independent [O/Fe] $\times$ [Fe/H] relation has been obtained for the galactic disk, from which 
we can write approximately

$${\rm [Fe/H]}  = \gamma  + \delta \ (\log {\rm O/H} + 12), \eqno(1)$$

\noindent
where $\gamma \simeq -10.841$  and $\delta \simeq 1.214$.
Here it has been implicitely assumed that log(O/H)$_\odot + 12 = 8.83$ (Grevesse \& Sauval 1998).
Note that Maciel et al. (2003) used a slightly larger value 
for the solar  neighbourhood, $\delta \simeq 1.4$, but the main conclusions of the present paper are not 
affected by this discrepancy. Since both O/H and [Fe/H] gradients are assumed to be linear, it 
is easy to see that

$$d{\rm [Fe/H]}/dR \simeq \delta \ d\log {\rm (O/H)}/dR, \eqno(2)$$

\noindent
which can be applied to the O/H gradients of Table 1. The derived [Fe/H] gradients are given 
in the last column of Table 1. 
\bigskip
\bigskip
%
\centerline{\psfig{figure=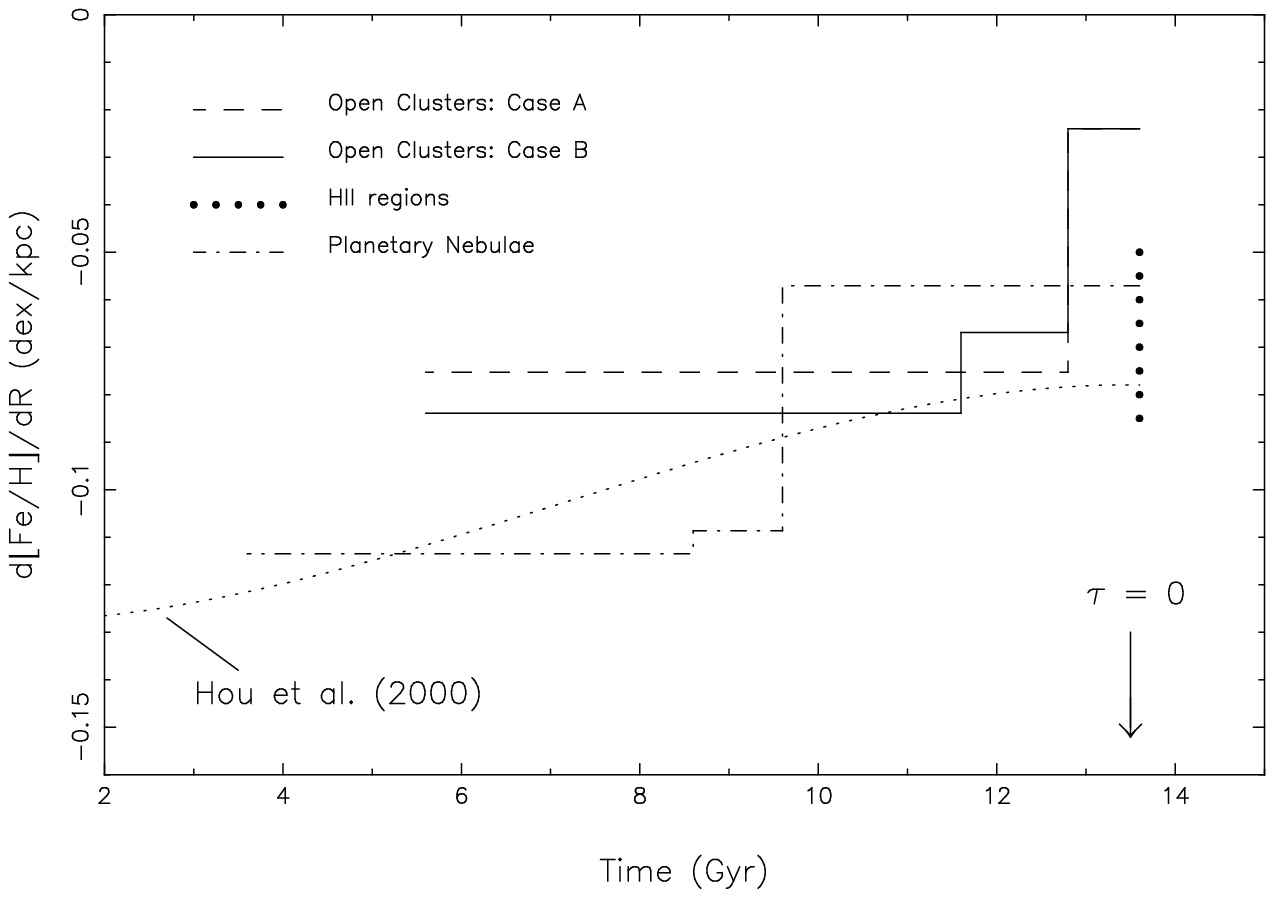,height=7.0 cm,angle=270} }
\bigskip\noindent
{\it Figure 1. Time variation of the [Fe/H] abundance gradient (dex/kpc).  The age of the galactic disk is 
assumed  to be 13.5 Gyr. The open cluster gradients are as follows:  Case A (dashed line) and Case B 
(solid line). The converted [Fe/H] gradients for the planetary  nebulae are shown by the dot-and-dashed 
lines, corresponding to Case B of Maciel et al. (2003). The solid dots at the present time ($\tau = 0$) show the 
average gradients of HII regions,  converted to [Fe/H] gradients as in the case of the planetary  nebulae. 
The dotted curve represents the time evolution of the [Fe/H] gradient according to the theoretical models 
of Hou et al. (2000).}
\bigskip\noindent
The results are better seen in Figure 1, where we plot the [Fe/H] gradients from the open 
clusters in Case A (dashed line) and Case B (solid line). The corrected planetary nebula 
[Fe/H] gradients (dot-and-dashed line) are shown for Case B only, which is more realistic, as 
discussed by Maciel et al. (2003). The average HII region gradients, converted to [Fe/H] 
gradients as in the case of the planetary nebulae, are shown as solid dots at $\tau \simeq 0$.
 
We have also 
included for comparison  purposes the time evolution of the [Fe/H] gradient predicted  by 
Hou et al. (2000), which is based on an inside-out  scenario for the formation of the disk using 
metallicity dependent yields (dotted curve).

It can be seen both from Figure 1 and Tables 1 and 2 that the open cluster gradients agree very 
well with the gradients derived from planetary nebulae, in the sense that both indicate a time 
flattening of the gradients during the lifetimes of these objects. The observational data also 
show a good agreement with theoretical models by Hou et al. (2000), as indicated by the 
dotted curve  in Figure 1. This conclusion holds in spite of the  relatively large uncertainty 
involved in the age determinations, especially in the case of the planetary nebula progenitor  
stars. Although the division of the open clusters into different  age groups is arbitrary, it can 
be seen that the time variation  of the gradients is not sensitive to the particular groups chosen. 
In fact, as long as the age groups contain a reasonably large  fraction of the total sample, the 
precise definition of the  groups is rather irrelevant. 

It can also be seen that the application of a correction factor to the O/H gradient as given  by 
the $\delta$ parameter given above leads to a better agreement between both sources of data. The 
flattening rate is  still uncertain, but from a comparison of both samples an average  rate of 
about 0.005 dex kpc$^{-1}$ Gyr$^{-1}$  can be obtained, which is similar to the earlier results by Maciel  
et al. (2003) and Chen et al. (2003). Although  this value is still uncertain, it clearly sets the 
order of  magnitude of the time flattening of the gradients, at least for  the time interval of 
roughly 10 Myr to 5 Gyr, for which our samples  are reasonably well defined. The gradient of 
about $-0.07$ dex/kpc derived for cepheid variables (Andrievsky et al. 2002)  also supports this 
conclusion, adopting ages in the range 20--300 Myr for these objects. A steeper rate at $\tau > 5$ 
Gyr cannot be ruled out, in view of the results for the planetary nebulae of group III. 
Regarding the most recent epochs,  $\tau < 10$ Myr, assuming that the HII region gradient  is 
representative of the present time, the planetary nebula data and the HII region results suggest 
a continuous flattening  in the last few Myr. Taking into account the open clusters of  Group I, 
there is apparently some steepening in the recent epoch, but it should be mentioned that most 
of the clusters in this group are located at $R < 10$ kpc, while the other groups are more evenly 
distributed in the galactic disk. Therefore, the derived gradient of Group I may in fact be a 
lower limit, which is supported by the lower correlation coefficient shown in Table 2. Clearly, 
more data is necessary to settle the fine points of this question.
\bigskip
{\it Acknowledgements. This work was partly supported by FAPESP and  CNPq.}
\bigskip
\bigskip\noindent
{\bf References}
\bigskip\noindent
\ref
[1]	Alib\'es, A., Labay, J., \& Canal, R., 2001, A\&A 370,1103
\ref
[2]	Andrievsky, S.M., Kovtyukh, V.V., Luck, R.E.,  L\'epine, J.R.D., Maciel, 
 	W.J., \& Beletsky, Yu. V. 2002 A\&A, 392, 491 
\ref
[3]	Chen, L., Hou, J. L., \& Wang, J. J. 2003, AJ 125, 1397
\ref
[4]	Chiappini, C., Matteucci, F., \& Romano, D. 2001, ApJ, 554, 1044
\ref
[5]	Costa, R. D. D., Uchida, M. M. M., \& Maciel, W. J. 2004,  A\&A (submitted)
\ref
[6]	Deharveng, L., Peña, M., Caplan, J., \& Costero,  R. 2000, MNRAS, 311, 329
\ref
[7]	Dias, W. S., Alessi, B. S., Moitinho, A., \& L\'epine, J. R. D. 2002, A\&A,  
     389, 871
\ref
[8]	Friel, E. D. 1995, An. Rev. A\&A, 33, 381
\ref
[9]	Friel, E. D. 1999, Ap\&SS, 265, 271
\ref
[10]	Friel, E. D., Janes, K. A., Tavarez, M., Scott, J., Katsanis, R., Lotz, J., Hong,
     L., \& Miller, N. 2002, AJ, 124, 2693
\ref
[11]	Grevesse, N. \& Sauval, A. J. 1998, Space Sci. Rev.,  85, 161
\ref
[12]	Henry, R. B. C. \&  Worthey, G. 1999, PASP, 111, 919
\ref
[13]	Hou, J. L., Prantzos, N., \& Boissier, S. 2000, A\&A, 362, 921
\ref
[14]	Maciel, W. J. 1997, in IAU Symp. 180, Planetary Nebulae,  ed. H. J. Habing,
     H. J. G. L. M. Lamers (Dordrecht: Kluwer), 397
\ref
[15]	 Maciel, W. J. 2000, in The Evolution of the Milky Way,  ed. F. Matteucci  
     \& F. Giovannelli (Dordrecht: Kluwer), 81
\ref
[16]	Maciel, W.J. 2002, Rev. Mex. AA - SC, 12, 207
\ref
[17]	Maciel, W. J. \& Costa, R. D. D. 2003, in Proc. IAU Symp. 209,   Planetary 
	Nebulae, ed. S. Kwok, M. Dopita \& R. Sutherland (San Francisco: ASP), 
     551
\ref
[18]	Maciel, W. J., Costa, R. D. D., \& Uchida, M. M. M. 2003, A\&A,  397, 667
\ref
[19] 	Maciel, W. J. \& K\"oppen, J. 1994, A\&A 282, 436
\ref
[20]	Maciel, W. J. \&  Quireza, C. 1999, A\&A, 345, 629
\ref
[21]	Phelps, R. 2000, in The Evolution of the Milky Way, ed. F. Matteucci \& 
     F. Giovannelli (Dordrecht: Kluwer), 239
\ref
[22]	Smartt, S. J. 2000, in The Evolution of the Milky Way, ed. F. Matteucci 
     \& F. Giovannelli (Dordrecht: Kluwer), 323
\ref
[23]	Tosi, M. 2003, in STScI Symposium, The Local Group as an  astrophysical 
	laboratory, astro-ph/0308463
\ref
[24]	V\'\i lchez, J.M. \&  Esteban, C. 1996, MNRAS, 280, 720
\ref

\bye